\documentclass[journal]{IEEEtran}
\IEEEoverridecommandlockouts
\usepackage{cite}
\usepackage{amsmath,amssymb,amsfonts}
\usepackage{algorithmic}
\usepackage{graphicx}
\usepackage{caption,subcaption}
\usepackage{booktabs}
\usepackage{adjustbox}
\usepackage[table,xcdraw]{xcolor}
\usepackage{textcomp}
\usepackage{multirow}
\usepackage{multicol}
\usepackage{booktabs}
\usepackage{bigstrut}
\usepackage{verbatim}
\usepackage{array}
\usepackage{rotating}
\usepackage{xcolor}
\usepackage{dblfloatfix}
\usepackage{soul}
\usepackage[a4paper, total={184mm,239mm}]{geometry}

\begin{document}

\title{Optimizing Phase-Scheduling with Throughput Trade-offs in AQFP Digital Circuits}
\author{Robert S. Aviles, Peter A. Beerel, \IEEEmembership{Senior Member, IEEE}

The authors are with the Department of Electrical and Computer Engineering,
University of Southern California, Los Angeles, CA 90007 USA (e-mail:
rsaviles@usc.edu; pabeerel@usc.edu)}

\markboth{
}{Robert S. Aviles, Peter A. Beerel: Optimizing Phase-Scheduling in AQFP Digital Circuits}

\maketitle

\begin{abstract}
Adiabatic Quantum-Flux-Parametron (AQFP) logic is a promising emerging superconducting technology for ultra-low power digital circuits, offering orders of magnitude lower power consumption than CMOS.  However, AQFP scalability is challenged by excessive buffer overhead due to path balancing technology constraints. 
Addressing this, recent AQFP works have proposed design solutions to reduce path balancing overhead using \textit{phase-skipping} and \textit{phase-alignment}.  Phase-skipping is a circuit-level technique that allows data transfer between AQFP gates clocked with non-consecutive clock phases. In contrast, phase-alignment is an architectural approach involving repeating input patterns to allow data transfer between AQFP gates across multiples of full clock cycles. While both techniques individually mitigate the area overhead of path-balancing, they have not yet been jointly explored. 
In this work, we present the first clock phase scheduling algorithm that combines phase-skipping and phase-alignment. We first present a minimum area method that on average, achieves a 25\% area reduction compared to phase-skipping alone and a 11\% reduction compared to phase-alignment.  We then extend the method to enforce a target throughput, enabling efficient area-performance trade-offs. With our throughput constrained optimization, we achieve on average 6.8\% area savings with a 2.62x increased throughput compared to the state-of-the-art phase-aligned method.

\end{abstract}

\begin{IEEEkeywords}
Superconducting logic circuits, design automation, circuits synthesis, beyond CMOS, digital circuits.
\end{IEEEkeywords}

\section{Introduction}

\IEEEPARstart{S}{uperconductive} digital electronics is a promising platform for high performance ultra-low power computing.  Superconducting electronic families have shown promise in reducing energy bottlenecks for a range of critical applications including neural network computing~\cite{DL_AQFP2019,BNN_AQFP2023}, post-quantum encryption, design of supercomputer microprocessors~\cite{MANA}, and enabling scalability of quantum computing~\cite{DigiQ,SFQ_ECC,SFQQAOA}. Superconducting logic achieves reduced power consumption through computing using Josephson Junctions (JJs), which operate based on the principles of quantum tunneling and achieve switching energies 1000x lower than CMOS.  

One of the most promising low power superconducting logic families is the adiabatic quantum flux parametron (AQFP)~\cite{AQFP} which operates adiabatically and consumes zero static power~\cite{zero_static} while achieving up to 5 GHz operation.  Even accounting for the energy overhead to cool to 4.2K, AQFP achieves two orders of magnitude advantage in EDP compared to state of the art semiconductor technologies \cite{cooling_overhead}.

While superconducting electronics offer compelling advantages in energy efficiency and speed, they impose unique technology constraints that hinder large-scale integration and demand custom EDA solutions~\cite{SFQ_EDA}.
\begin{figure}[t]
\includegraphics[width=0.9\columnwidth]{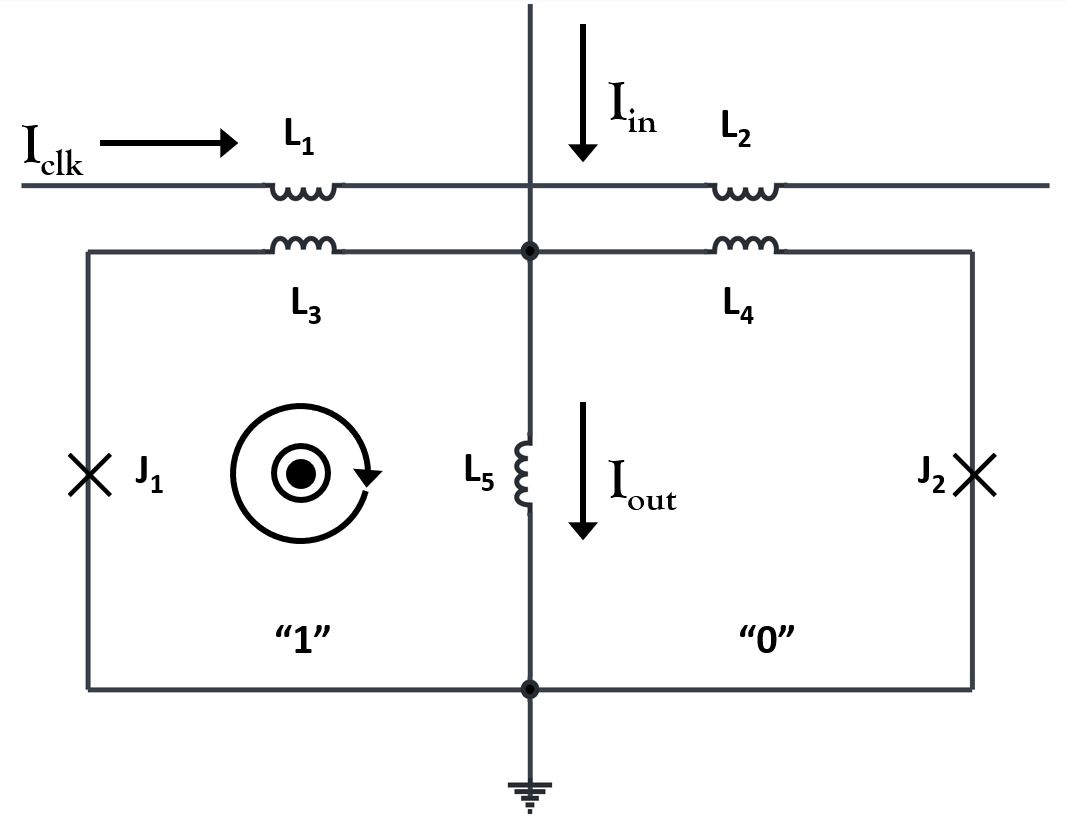}
\centering
\caption{Circuit Construction of AQFP Buffer.  Current directions shown for when flux is in the left loop, corresponding to a logical "1".}\label{fig:AQFPBuffer}
\label{buf}
\end{figure}
 In AQFP logic specifically, every gate and splitter must be clocked, and data transfers require carefully timed overlapping AC clock phases. These strict timing requirements lead to significant area overhead due to the insertion of clocked buffers to equalize path delays.  In some designs, these buffers can consume up to 90\% of the total circuit area~\cite{buffer_costs}.
To address this challenge, prior work has proposed various methods for efficient path balancing by optimizing buffer and splitter  insertion~\cite{dp_OG,bs-2,heuristicASP,DAC22,SOTA}.   More recently, researchers have explored techniques to relax strict path balancing constraints to reduce area overhead. 

 Phase-skipping increases the number of usable clock phases, allowing imbalances to be resolved at the circuit level by exploiting more flexible timing constraints~\cite{Nphaseclk,AvilesNPhase}. In contrast, phase-alignment takes an architectural approach: it reuses input patterns across cycles to resolve imbalances that are full clock-cycle multiples—trading off throughput for area reduction~\cite{phaseMatch}. 

While both phase-skipping and phase-alignment have independently demonstrated area benefits—and are known to be compatible~\cite{phaseMatch}—no prior work has jointly optimized them within a unified framework.

\begin{figure*}[htbp]
\includegraphics[width=0.95\linewidth]{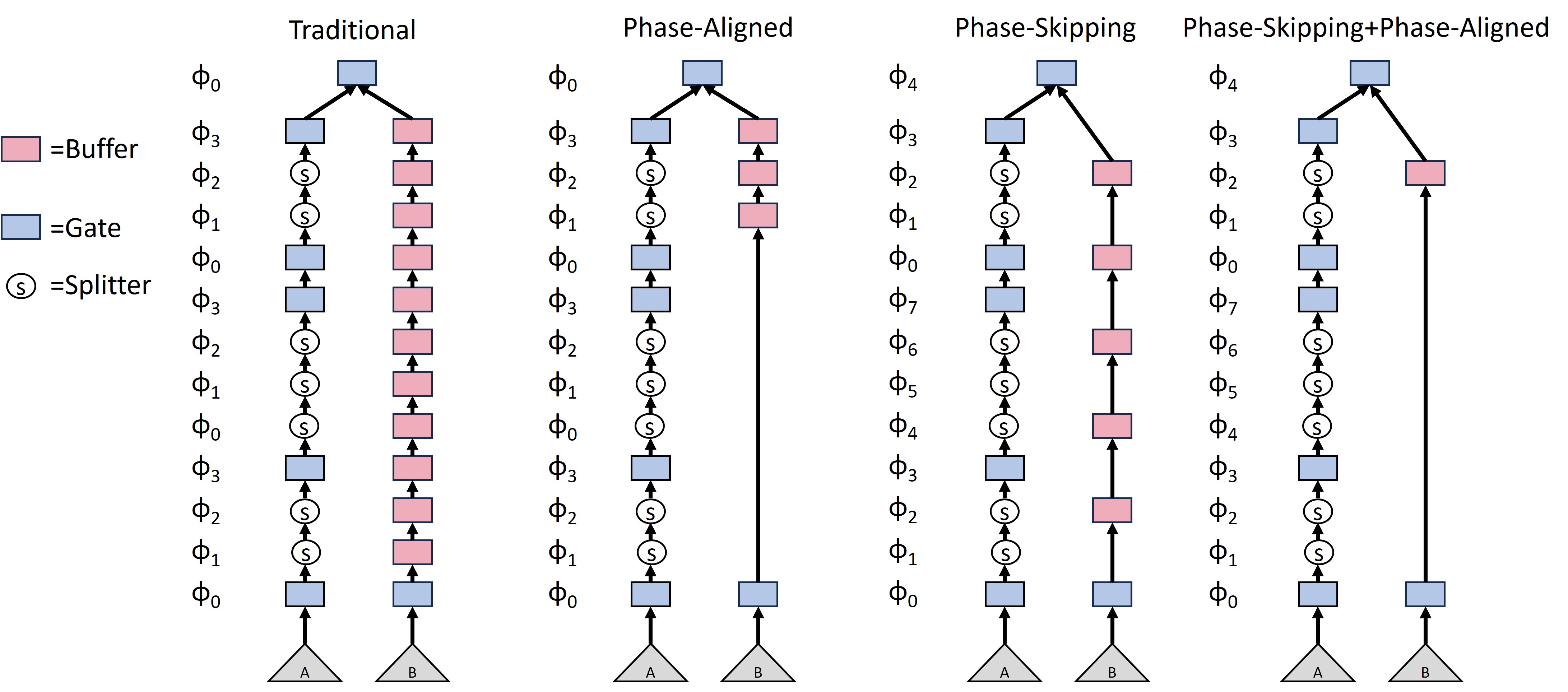}
\centering
\caption{Buffer insertion along highly imbalanced paths for various clocking schemes (off path fan-ins/fanouts omitted for clarity).  Imbalances in logical gates are often magnified by the insertion of clocked splitters to achieve fanout.}
\label{fig:imbalanced_4schemes}
\end{figure*}

This paper formulates an integer linear program (ILP) framework to optimize clock phase assignments and buffer insertion that jointly leverages both phase-skipping and phase-alignment.
\begin{itemize}
    \item We first show how our ILP framework can be formulated to minimize buffer insertion achieving a \textbf{25\%} area reduction over phase-skipping alone and \textbf{11\%} area reduction over phase-alignment alone. 
    \item We then show how our ILP can be extended to be throughput-aware, enforcing a target throughput in phase-aligned circuits. With this formulation, we achieve on average \textbf{2.62×} higher throughput and 6.7\% less area than the state-of-the-art phase-aligned method.
\end{itemize}

\section{Background}
\subsection{AQFP Logic}

AQFP logic is based on a 
universal Majority-3 based logic family
constructed from four fundamental cells: the AQFP buffer, branch cells for merging and splitting signals, NOT cells, and constant sources (for fixed logical 0 or 1)~\cite{scl}. In particular, a 3-input merge cell outputs a value based on the sum of its input currents~\cite{scl,maj3} driven by buffers, constants, or NOT gates (which are buffers with inverted outputs). To support fanout greater than one, a branch cell is combined with a buffer, forming a clocked splitter. 

As shown in Fig.~\ref{fig:AQFPBuffer}, an AQFP buffer encodes logic in the direction of its output current by storing a single flux quantum in one of two superconducting loops. The flux location determines the output polarity: positive current flow represents a logical ‘1’, and negative flow a logical ‘0’.

A central design challenge in AQFP is satisfying a buffer's strict timing constraints. To reliably store flux, the input signal must arrive before the rise of the AC excitation current ($I_{clk}$), remain valid for the required setup time, and continue to hold through the buffer's hold period~\cite{Nphaseclk}. Moreover, the buffer only produces output during the active portion of its clock, effectively tying data validity to a clock phase. To ensure proper operation, connected gates must be assigned clock phases that both overlap and satisfy these setup/hold constraints~\cite{overlap}.

Because buffers are the basic building block of all AQFP logic gates and splitters--every element must be explicitly clocked and aligned with strict timing. Traditionally, data synchronization has been ensured using 3-phase or 4-phase clocking, where each clock has the same frequency with shifted phase and all gates have fully balanced logic levels\footnote{In modern placement flows, logic level and clock phase are tightly coupled: all gates at the same level are placed in the same row and share the same clock. Thus, level assignment and phase assignment are effectively interchangeable.}.  While this design is simple, it leads to significant path balancing overhead. 

\begin{figure}[htbp]
\includegraphics[width=\columnwidth]{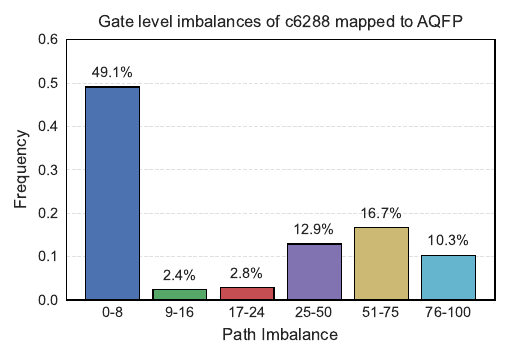}
\centering
\caption{When c6288 is mapped to AQFP technology with balanced splitter trees inserted, over 50\% of logic gates have an imbalance in inputs exceeding 9 levels.  The average imbalance is 32.6 levels.  }\label{fig:Imbalance_dist}
\end{figure}
\subsection{AQFP Clocking Schemes}

To reduce AQFP’s area overhead, alternative clocking schemes have been proposed that achieve data synchronization without requiring fully balanced logic levels. \textit{Phase-skipping}, introduced in~\cite{Nphaseclk}, leverages the fast switching speed of AQFP circuits and reduces the delay between adjacent clock phases and increasing the number of phases in a clock cycle without increasing its period. This leads to each phase overlapping with more neighbors, enabling data transfer across paths with limited path imbalances. When compared to path balanced clocking, clock phase assignment optimization using phase-skipping has demonstrated average reductions in Josephson Junction (JJ) count of 26\% with a 1-phase skip and up to 40\% with 4-phase skips~\cite{AvilesNPhase}.

Separately, \textit{phase-alignment} was proposed in~\cite{phaseMatch} as an architectural strategy that relaxes path balancing requirements by repeatedly applying input values, enabling data alignment between gates with different clock cycle depths. This scheme is implemented using Non-Destructive Readout Cells (NDROs) which also provide for increased fanout capacity from the primary inputs~\cite{phaseMatch}.  Since data must still be transmitted only between adjacent phase assignments, the imbalances must equal an integer number of full clock cycles without buffer insertion. We refer to this mechanism as a \textit{cycle-skip}, distinguishing it from phase-skipping across overlapping but non-consecutive phases. For example, in an 8-phase system, a cycle-skip of $S$ allows imbalances of $8 \cdot S$ logic levels to be absorbed without additional buffering.

Phase-alignment offers substantial area savings, reducing JJ count by an average of 45\% compared to path balanced circuits ~\cite{phaseMatch}. However, this benefit comes at the cost of throughput, as input repetition inhibits pipelining. Accordingly, throughput degrades proportionally with the number of repetitions, introducing a key trade-off between area and performance. While previous work proposes limiting the maximum phase-skip to mitigate degradation, we introduce a more precise approach for throughput management in Section~\ref{throughput}.

\section{Combining Phase-Skipping and Phase-Alignment}

We propose to unify phase-skipping and phase-alignment within a single optimization framework. We formulate an integer linear program (ILP) that jointly assigns clock phases and minimizes buffer insertions, subject to user-defined constraints on both phase-skipping and throughput. As illustrated in Figure~\ref{fig:imbalanced_4schemes}, our combined method can resolve a logic path imbalance of 11 levels using only one buffer—compared to 11 buffers with traditional methods. Such highly imbalanced paths, often introduced by clocked splitters used for fanout, are prevalent in AQFP circuits, as shown in Fig.~\ref{fig:Imbalance_dist}.

\subsection{Enhanced Management of Imbalances}

Phase-skipping is a circuit-level technique that involves increasing the number of overlapping clock phases to allow data transfer across $P$ non-consecutive phases. For example, in an N-phase clocking scheme $P = N/4$ phases overlap ~\cite{Nphaseclk}. Let $\Delta L_{ij}$ be the level imbalance between gates $i$ and $j$. Then, the buffer cost is $\lfloor \frac{(\Delta L_{ij} - 1)}{P} \rfloor$ when using phase-skipping alone.

In contrast, phase-alignment acts requires architectural-level changes involving repeating inputs to a block to tolerate logic-level imbalances that span one or more full clock cycles. More precisely, phase-alignment can resolve level differences that are exact multiples of \textit{N} with zero buffer cost as shown in Fig.~\ref{fig:imbalanced_4schemes}. Any remaining imbalance incurs a buffer cost of ($\Delta L_{ij}-1$ ) \textit{mod N}, 

\begin{figure}[htbp]
\includegraphics[width=\columnwidth]{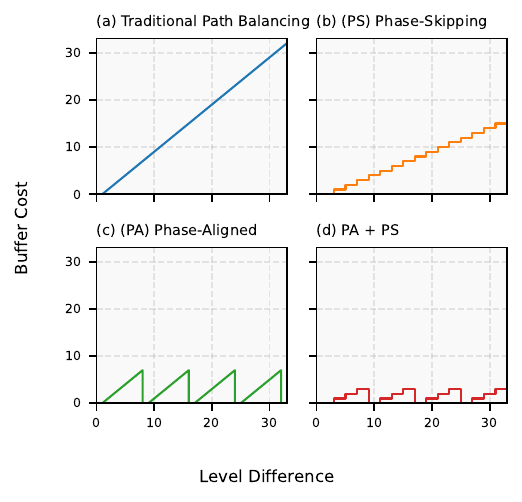}
\centering
\caption{Buffer insertion costs for each clocking scheme based on level imbalances. (Phase alignment is shown with 8-phases highlight features in comparison to 8-phase phase-skipping circuits)}
\label{fig:scheme_costs}
\end{figure}

When both techniques are used in tandem, phase-alignment absorbs full cycles of imbalance and phase-skipping resolves the remaining offset. This yields the combined buffer cost:
\[
\left\lfloor \frac{(\Delta L_{ij} - 1) \bmod N}{P} \right\rfloor
\]

Figure~\ref{fig:scheme_costs} illustrates these cost functions for $N = 8$. The combination of phase-skipping and phase-alignment results in a nonlinear and non-monotonic cost profile: even small changes in $\Delta L_{ij}$ can lead to different combinations of cycle-skips and phase-skips, resulting in sharply varying buffer requirements.

\begin{figure}[htbp]
\includegraphics[width=0.9\columnwidth]{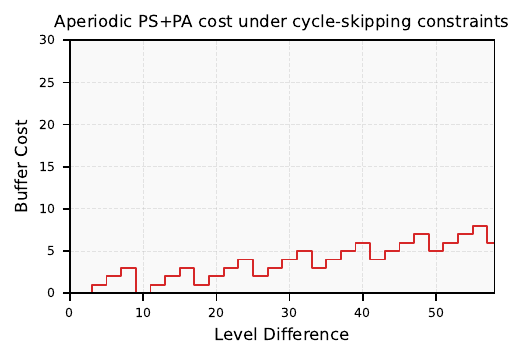}
\caption{Buffer insertion costs become aperiodic when cycle-skipping constraints are enforced.   When $S_{max} = 1$, buffers must be inserted to limit each connection to span only 1 cycle.}
\label{fig:scheme_aperiodic}
\end{figure}

Practical place-and-route constraints introduce additional complexity. To avoid excessive wire lengths, the total level difference between connected gates may need to be bounded. We model this with a maximum cycle-skipping constraint, denoted $S_\text{max}$. This constraint breaks the natural periodicity of phase-alignment, as shown in Fig.~\ref{fig:scheme_aperiodic} for $S_\text{max} = 1$, and must be accounted for during clock scheduling.

\begin{figure*}[t!]
\includegraphics[width=\linewidth]{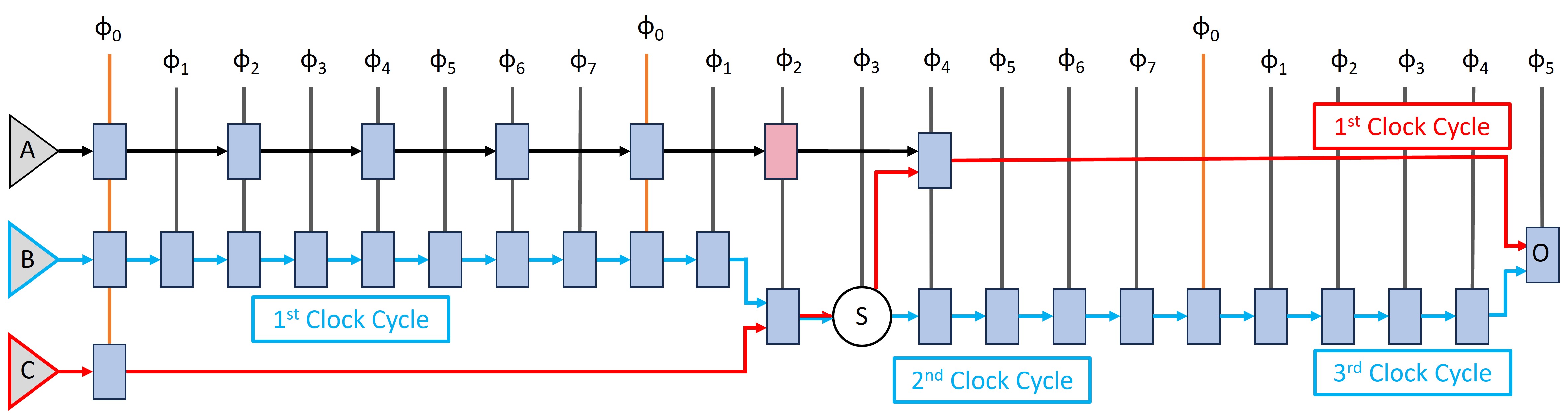}
\centering
\caption{Despite the maximum skip between connections being limited to 1 cycle, an accumulated imbalance of 2 clock cycles requires inputs to be presented 3 times resulting in a 1/3 maximum throughput.}
\label{fig:Throughput}
\end{figure*}
\subsection{ILP Formulation}\label{ilp}

We formulate an ILP based optimization algorithm for clock phase assignment that minimizes total buffer insertion while satisfying AQFP technology constraints.

The input to the algorithm is an AQFP logic netlist, modeled as a directed acyclic graph (DAG) $G = (V, E)$, where each internal node $v_i \in V$ represents a logic gate or splitter, and each directed edge $E_{ij} \in E$ corresponds to a signal connection from node $i$ to node $j$. Circuit inputs and outputs are modeled as source and sink nodes, respectively. Each node is assigned an integer level $L_i$, which directly determines its clock phase assignment.

Our objective is to minimize the total number of inserted buffers. For each edge $(i,j)$, we define the cost $C_{ij}$ as the number of buffers needed to legalize the level difference between $L_i$ and $L_j$. 

To manage the complex cost behavior that arises from jointly using phase-skipping and cycle-skipping, buffer insertion costs are split into two distinct and independent components,
$\alpha_{ij}$ and 
$\beta_{ij}$.

This decomposition reflects the different mechanisms of phase-skipping and cycle-skipping, simplifying an otherwise irregular and piecewise cost function. Specifically:
\begin{itemize}
    \item $\alpha_{ij}$ models phase-skipping buffers, each of which can compensate for $P$ levels of imbalance via overlapping clock phases.
    \item $\beta_{ij}$ models cycle-skipping buffers, where each unit compensates for $1 + S_{\text{max}} \cdot N$ levels of imbalance aligned to full clock cycles.
\end{itemize}

By defining these two variables, our optimization framework can reason separately about short-range and long-range imbalances, allowing more efficient search over the level assignment space. This modeling strategy is key to simplifying ILP-based formulation.

The optimization is subject to the following constraints:
\begin{itemize}
    \item \textbf{Level bounds:} The level difference between connected nodes must be consistent with the number of buffers and cycle-skips assigned:
    \setcounter{equation}{1}
    \begin{equation}\label{eq:cij_cost1}
 1+S_{ij}\cdot N + \beta_{ij}\leq L_j-L_i  \hspace*{0.6in}  
\end{equation}

    \item \textbf{Upper bound:} To ensure timing feasibility, level differences must not exceed the maximum imbalance tolerated by the inserted buffers:
\begin{equation}\label{eq:cij_cost2}
  L_j-L_i \leq {(\alpha_{ij}+1)\cdot P +S_{ij}\cdot N + \beta_{ij}} 
\end{equation}
    
    \item \textbf{Cycle-skipping limit:} $S_{ij}$ denotes the number of full clock cycle skips. This constraint ties cycle-skipping decisions to buffer count and enforces the allowed skip range:
\begin{equation}\label{eq:Sij_limit}
 \beta_{ij} \cdot S_{max} \leq S_{ij} \leq  (1+\beta_{ij}) \cdot S_{max}  
\end{equation}
\end{itemize}

Additionally, to match phase-alignment requirements from \cite{phaseMatch}, primary input nodes are allowed to occupy levels 3 through 5. This corresponds to the permissible range for repeated inputs driven by D-latch cells. For uniform clocking and signal arrival times, all primary outputs are constrained to share a common level.
Our optimization problem becomes: 
\setcounter{equation}{0}
\begin{equation} \label{eq:newcost}
    \text{Minimize:} \sum_{E_{ij}=1}{C_{ij}}
\end{equation}
\begin{equation*}
    \text{subject to:}
\end{equation*}
\begin{equation}
 1+S_{ij}\cdot N + \beta_{ij}\leq L_j-L_i  \hspace*{0.8in}  \forall (i,j) \in E_{i,j},
\end{equation}
\begin{equation}
  L_j-L_i \leq {(\alpha_{ij}+1)\cdot P +S_{ij}\cdot N + \beta_{ij}} \hspace*{0.1in}  \forall (i,j) \in E_{i,j},
\end{equation}
\begin{equation}
 \beta_{ij} \cdot S_{max} \leq S_{ij} \leq  (1+\beta_{ij}) \cdot S_{max}  \hspace*{0.4in}  \forall (i,j) \in E_{i,j},
\end{equation}
\begin{equation}
C_{ij} = \alpha_{ij} + \beta_{ij}  \hspace*{1.55in}  \forall (i,j) \in E_{i,j},
\end{equation}
\begin{equation}
 L_i = L_{outputs}  \hspace*{1.95in} \forall i \in O, 
\end{equation}
\begin{equation}
   3 \leq L_i \leq 5    \hspace*{2.15in} \forall i \in I,
\end{equation}
\begin{equation}
\label{eq:Integral}
    L_i, C_{ij},\alpha_{ij},\beta_{ij},S_{ij} \in \mathbb{N} \hspace*{1.05in}  \forall (i,j) \in E_{i,j}
\end{equation}

Solving this optimization problem yields level assignments $L_i$ for all nodes and directs where buffers must be inserted along each edge.  The resulting circuit can then be passed to the place-and-route stage of an AQFP design flow.

\section{Managing Throughput Trade-offs}\label{throughput}
\subsection{Throughput Trade-offs in a Phase-Aligned System}

While gate-level clocking in AQFP requires data synchronization constraints, it also produces deeply pipelined circuits that affect architectural-level throughput. Our formulation in Section \ref{ilp} focused solely on minimizing area under AQFP timing constraints. In this section, we introduce architectural level constraints to manage throughput.

AQFP circuits are inherently pipelined. In a fully path-balanced design, new inputs can be applied on every clock cycle, enabling a maximum throughput equal to the clock frequency (e.g., 5 GHz for a 5 GHz clock). However, phase-alignment techniques require the same input to be reapplied over multiple clock cycles. This repetition reduces effective throughput proportionally to the number of repeated input cycles~\cite{phaseMatch}.

A conservative upper bound on the number of repetitions is 
$\lceil\frac{D}{N}\rceil$, where $D$ is the circuit depth and $N$ is the number of clock phases. However, a finer-grained analysis of path structure can enable more aggressive throughput optimization even in deep circuits. 

While prior work~\cite{phaseMatch} mentions limiting maximum phase-skips to control throughput degradation, their approach does not generalize to arbitrary circuit structures. In particular, we show that it is the cumulative cycle-skipping across a path, not a per-edge limit discussed in \cite{phaseMatch}, that determines the worst-case input repetition.

Figure~\ref{fig:Throughput} illustrates this point. Although each edge in the circuit uses at most one cycle-skip (i.e., 8 levels in an 8-phase clock), the path from input \textit{C} to gate \textit{O} contains 2 cycle-skips and arrives in cycle 1, while the path from input \textit{B} to \textit{O} arrives in cycle 3.  Since AQFP cells are stateless, all inputs to a gate must be presented simultaneously, requiring the earlier signal to be reapplied for two additional cycles.

Therefore, maintaining a desired throughput requires careful control over the total number of accumulated cycle-skips on each path. Our framework explicitly models this constraint, enabling designers to trade off throughput and area in a principled way.

\subsection{Target Throughput Constraints}

To constrain throughput degradation in phase-aligned AQFP circuits, we introduce a new variable $R_i$ for each node $i$ in the DAG. This variable captures the number of input repetitions required to ensure proper data alignment at node $i$, accounting for all accumulated cycle-skips along the longest path from any primary input (PI) to $i$. 

The maximum value across all $R_i$ variables in the graph defines the overall input repetition requirement of the circuit.  Consequently, the circuits throughput is reduced by a factor of $\frac{1}{R_{max}+1}$, since each input must be applied for $R_{max} + 1$ cycles. 

To enforce this throughput constraint, we augment our ILP formulation with two new constraints:
\begin{itemize}
    \item (Eq.~\ref{eq:Ri_formulation}) that ensures each $R_j$ reflects the maximum of all upstream values $R_i$ plus any new cycle-skipping incurred along edge (i,j)
    \begin{equation}\label{eq:Ri_formulation}
 R_i + S_{ij} \leq R_j \hspace*{0.3in}  \forall (i,j) \in E_{i,j},
\end{equation}
    \item (Eq.~\ref{eq:Ri_Rmax}) bounding the total number of input repetitions by user-specified $R_{max}$ and ensuring performance targets are met.
    \begin{equation}\label{eq:Ri_Rmax}
 R_i \leq R_{max} \hspace*{0.3in}  \forall (i) \in V
\end{equation}
\end{itemize}

These added constraints enable optimization to minimize area subject to architectural throughput bounds, effectively co-optimizing for performance and area.


\begin{table*}[htbp]
  \centering
  \caption{Comparison of Buffer and Splitter Insertion Methods on AQFP Benchmarks}
  \label{tab:area_comparison}
  \renewcommand{\arraystretch}{1.1}
  \begin{tabular}{lcc cc ccc ccccc}
    \toprule
    \multirow{2}{*}{\textbf{Benchmark}} 
    & \multicolumn{2}{c}{Original Netlist} 
    & \multicolumn{2}{c}{SOTA 1-Phase Skipping \cite{AvilesNPhase}} 
    & \multicolumn{3}{c}{SOTA Phase Alignment \cite{phaseMatch}} 
    & \multicolumn{5}{c}{Our Work (Phase Alignment + 1P Skipping)} 
    \\
    \cmidrule(lr){2-3}
    \cmidrule(lr){4-5}
    \cmidrule(lr){6-8}
    \cmidrule(lr){9-13}
    & Gates & JJs 
    & BS & JJs 
    & BS & JJs & MPS 
    & BS & JJs & MPS & \% vs PS & \% vs PA  
    \\
    \midrule
    \texttt{adder1} & 7 & 42 & 8 & 58 & 5 & 52 & 4 & 3 & 48 & 8 & 17.2\% & 7.6\%  \\
    \texttt{adder8} & 77 & 462 & 168 & 798 & 87 & 636 & 24 & 57 & 576 & 25 & 27.8\% & 9.4\%  \\
    \texttt{mult8} & 439 & 2634 & 957 & 4548 & 681 & 3996 & 60 & 418 & 3470 & 56 & 23.7\% & 13.1\%  \\
    \texttt{counter16} & 29 & 174 & 39 & 252 & 52 & 278 & 20 & 27 & 228 & 9 & 9.5\% & 18\%  \\
    \texttt{counter32} & 82 & 492 & 107 & 706 & 131 & 754 & 28 & 71 & 634 & 17 & 10.2\% & 15.9\%  \\
    \texttt{counter64} & 195 & 1170 & 226 & 1622 & 309 & 1788 & 36 & 169 & 1508 & 17 & 7\% & 15.7\%  \\
    \texttt{counter128} & 428 & 2568 & 484 & 3536 & 656 & 3880 & 44 & 355 & 3278 & 17 & 7.3\% & 15.5\%  \\
    \texttt{c17} & 6 & 36 & 3 & 42 & 5 & 46 & 0 & 1 & 38 & 1 & 9.5\% & 17.3\%  \\
    \texttt{c432} & 121 & 726 & 440 & 1606 & 147 & 1020 & 28 & 78 & 882 & 25 & 45.1\% & 13.5\%  \\
    \texttt{c499} & 387 & 2322 & 1247 & 4816 & 407 & 3136 & 24 & 303 & 2928 & 25 & 39.2\% & 6.6\%  \\
    \texttt{c880} & 306 & 1836 & 798 & 3432 & 516 & 2868 & 40 & 194 & 2224 & 41 & 35.2\% & 22.5\%  \\
    \texttt{c1355} & 389 & 2334 & 1227 & 4788 & 398 & 3130 & 24 & 296 & 2926 & 25 & 38.9\% & 6.5\%  \\
    \texttt{c1908} & 289 & 1734 & 1030 & 3794 & 364 & 2462 & 28 & 213 & 2160 & 25 & 43.1\% & 12.3\%  \\
    \texttt{c2670} & 368 & 2208 & 969 & 4146 & 351 & 2910 & 32 & 226 & 2660 & 33 & 35.9\% & 8.6\%  \\
    \texttt{c3540} & 794 & 4764 & 1312 & 7388 & 1060 & 6884 & 44 & 664 & 6092 & 41 & 17.5\% & 11.5\%  \\
    \texttt{c5315} & 1302 & 7812 & 3200 & 14212 & 1337 & 10486 & 40 & 821 & 9454 & 41 & 33.5\% & 9.8\%  \\
    \texttt{c6288} & 1870 & 11220 & 7485 & 26190 & 3206 & 17632 & 168 & 2013 & 15246 & 169 & 41.8\% & 13.5\%  \\
    \texttt{c7552} & 1394 & 8364 & 4489 & 17342 & 1860 & 12084 & 56 & 1238 & 10840 & 57 & 37.5\% & 10.2\%  \\
    \texttt{sorter32} & 480 & 2880 & 480 & 3840 & 448 & 3776 & 0 & 448 & 3776 & 1 & 1.6\% & 0\%  \\
    \texttt{sorter48} & 880 & 5280 & 720 & 6720 & 960 & 7200 & 0 & 720 & 6720 & 1 & 0\% & 6.6\% \\
    \texttt{alu32} & 1513 & 9078 & 9711 & 28500 & 1969 & 13016 & 156 & 1422 & 11922 & 153 & 58.2\% & 8.4\% \\
    \midrule
    \rowcolor[rgb]{.9,.9,.9}
    \textbf{Total} & -- & -- & 35100 & 138336 & 14949 & 98034 & -- & 9737 & 87610 & \textbf{Avg} & \textbf{25.7\%} & \textbf{11.6\%} \\
    \bottomrule
  \end{tabular}
\end{table*}

\begin{table}[htbp]
  \centering
  \caption{Throughput Comparison of Phase-Alignment vs Our Work on AQFP Benchmarks}
  \label{tab:throughput_comparison}
  \renewcommand{\arraystretch}{1.1}
  \begin{tabular}{l cr ccc c}
    \toprule
    \multirow{2}{*}{\textbf{Benchmark}} 
    & \multicolumn{2}{c}{SOTA PA~\cite{phaseMatch}} 
    & \multicolumn{3}{c}{Our Work} 
    & \multirow{2}{*}{\shortstack{\textbf{Throughput}\\\textbf{Ratio}}}  \\
    
    \cmidrule(lr){2-3}
    \cmidrule(lr){4-6}
    
    & JJs &  T 
    & JJs & T & \% JJ 
    &  \\

    \midrule
    \texttt{adder1}       & 52  & 1/2       & 48    & 1/2    & 7.7\% & 1 \\
    \texttt{adder8}       & 636  & 1/7      & 604   & 1/3    & 5\% & 2.33 \\
    \texttt{mult8}       & 3996 & 1/16    & 3920  & 1/4    & 1.9\%  & 4 \\
    \texttt{counter16}        & 278   & 1/6      & 236   & 1/2    & 15.1\% & 3 \\
    \texttt{counter32}      & 754   & 1/8      & 650   & 1/2    & 13.8\% &  4 \\
    \texttt{counter64}     & 1788  & 1/10    & 1560  & 1/2    & 12.8\% &  5 \\
    \texttt{counter128}    & 3880  & 1/12    & 3392  & 1/2    & 12.6\% &  6 \\
    \texttt{c17}            & 46    & 1        & 38    & 1      & 17.4\%   &  1 \\
    \texttt{c432}           & 1020 & 1/8     & 928   & 1/4    & 9\%   & 2 \\
    \texttt{c499}           & 3136  & 1/7     & 3052  & 1/4    & 2.7\%   & 1.75 \\
    \texttt{c880}          & 2868  & 1/11    & 2732  & 1/3    & 4.7\%  & 3.66 \\
    \texttt{c1355}          & 3130 & 1/7     & 3020  & 1/4    & 3.5\%   & 1.75 \\
    \texttt{c1908}        & 2462  & 1/8     & 2308  & 1/4    & 6.3\%   & 2 \\
    \texttt{c2670}         & 2910   & 1/9     & 2760  & 1/4    & 5.2\% &  2.25 \\
    \texttt{c3540}        & 6884  & 1/12    & 6630  & 1/3    & 3.7\%   & 4 \\
    \texttt{c5315}        & 10486  & 1/11    & 9958 & 1/5    & 5\% &  2.2 \\
    \texttt{c6288}        & 17632  & 1/43    & 17470 & 1/14   & 0.9\% &  3.31 \\
    \texttt{c7552}         & 12084  & 1/15    & 11356 & 1/8    & 6\% &  1.88 \\
    \texttt{sorter32}       & 3776  & 1       & 3776  & 1      & 0\%  & 1 \\
    \texttt{sorter48}      & 7200  & 1       & 6720  & 1      & 6.7\%   &  1 \\
    \texttt{alu32}         & 13016  & 1/40    & 12710 & 1/20   & 2.4\% &  2 \\
    \midrule
    \rowcolor[rgb]{.9,.9,.9}
    \textbf{Total} &  98034 & --  & 93868 & \textbf{Avg} & \textbf{6.8\%} & \textbf{2.62} \\
    \bottomrule
  \end{tabular}
\end{table}

\section{Experimental Results}
\subsection{Results}

All circuit comparisons are performed using common netlists acquired from \cite{benchmarks}.  For our work, we insert balanced splitter trees, leaving any optimization of splitter tree topologies for joint phase-alignment and phase-skipping as future work.  To ensure consistency, we adopt the same experimental constraints as \cite{phaseMatch}: a maximum splitter fanout of 3, PI's with a fanout of 2, and PI levels constrained to 3, 4, or 5.  

Table~\ref{tab:area_comparison} presents area comparisons between our method and two state-of-the-art approaches: phase-skipping circuits from \cite{AvilesNPhase} and phase-aligned circuits from \cite{phaseMatch}. In this set of experiments, we focus purely on area optimization without applying throughput constraints. We also report the total number of buffers and splitters (BS) and the maximum phase skip (MPS) used in each circuit, consistent with the evaluation in \cite{phaseMatch}.

Because both baseline 
methods (\cite{AvilesNPhase} and \cite{phaseMatch}) are based on polynomial-time algorithms, we enforce a fair runtime cap on our ILP formulation by limiting it to 10 minutes per circuit. The best feasible solution found within this time limit is reported, regardless of whether the solver reached optimality. Even under these time constraints, our method achieves an average of 25\% area reduction compared to phase-skipping and 11\% improvement over phase-aligned solutions.

In Table~\ref{tab:throughput_comparison}, we evaluate the impact of incorporating throughput constraints into our optimization. The authors of \cite{phaseMatch} propose that limiting the maximum phase skip (MPS) can serve as a mechanism to control throughput in phase-aligned circuits. We adopt this assumption and convert their reported MPS values into throughput. Given that their implementation uses a 4-phase clock, throughput can be calculated as $\frac{1}{\frac{MPS}{4} +1}$, where the denominator represents the number of cycles over which each input must be repeated. 

For our method, we directly incorporate throughput as a constraint by limiting accumulated cycle-skips and set the throughput target for each circuit to the highest value that still achieves area equal to or better than the baseline. Under these conditions, our method demonstrates an average 2.62× improvement in throughput compared to phase alignment~\cite{phaseMatch}, while also reducing area by 6.7\% on average, highlighting the effectiveness of optimizing for area and throughput jointly.

\subsection{Discussion}

Beyond these results, the path toward larger systems highlights scalability as a key opportunity. While optimal convergence of ILPs is not tractable for very large circuits, they often reach high-quality feasible solutions well before full optimality, making them practical for design exploration and benchmarking. Moreover, the formulation itself provides a principled foundation from which scalable heuristics and approximation methods can be derived. In this way, our ILP serves not only as a solver but also as a blueprint for future algorithmic advances in AQFP optimization.

As of yet, no AQFP cell placement flow has been developed to support circuits that exploit phase-skipping or phase-alignment, limiting the immediate deployment of such clocking optimizations. Nonetheless, the significant area and throughput benefits demonstrated by our joint optimization framework—an additional 11\% area reduction or 2.6× throughput improvement—underscore the value of these techniques and highlight their promise for future AQFP system design.

Current AQFP design flows rely heavily on row-based placement, where each row is clocked by a single AC phase \cite{AQFPPlacement}. This layout paradigm poses challenges for imbalanced clocking schemes, as connected gates assigned to different phases must be placed in distant rows—potentially violating wirelength or timing constraints. However, this row assignment restriction is a design convention, not a fundamental limitation of AQFP hardware.

As fabrication processes and physical design methodologies mature, non-row-based layouts such as circuit folding—where cells with different clock phases are interleaved in placement—could become viable. These methods would directly support the flexible clock phase assignments generated by our ILP-based optimization framework. In that context, phase-skipping and phase-alignment become powerful, deployable tools for improving area in deeply pipelined AQFP systems.

\section{Conclusions}
In this work, we addressed the unification of two area-reduction techniques for AQFP logic: phase-skipping and phase-assignment.  We first described how these techniques efficiently manage the substantial path imbalances present in AQFP circuits. We then introduced a novel integer linear programming (ILP) formulation capable of handling the nonlinear and aperiodic cost structure that emerges when combining these two clocking approaches.

Using this formulation, we demonstrated an average of 25\% area reduction compared to 1-phase skipping, and a 11\% improvement over phase-alignment techniques. Beyond area, we examined the throughput implications of phase-alignment, emphasizing the need to minimize accumulated cycle-skips to maintain performance. By extending our ILP to include throughput constraints, we achieved an average 2.62× improvement in throughput compared to prior phase-alignment methods.

These results motivate future research in imbalanced AQFP circuit design in two key directions: (1) the development of physical design flows that support non-uniform clock phase placement, such as row folding, and (2) the creation of scalable heuristic or approximate solvers to extend the applicability of our optimization framework to very large circuits.

Together, these advances help expand the range of design strategies for unlocking AQFP’s energy efficiency in high-performance, area-constrained applications.

\bibliographystyle{IEEEtran}
\bibliography{bibliography}

\end{document}